\documentclass[11pt]{article}

\usepackage{a4wide}
\usepackage[sc]{mathpazo}
\usepackage{amsmath}
\usepackage{amssymb}
\usepackage{amsthm}
\usepackage{lmodern}

\usepackage{enumitem}
\usepackage{color}

\usepackage{stmaryrd}

\DeclareFontFamily{U}{mathx}{\hyphenchar\font45}
\DeclareFontShape{U}{mathx}{m}{n}{
      <5> <6> <7> <8> <9> <10>
      <10.95> <12> <14.4> <17.28> <20.74> <24.88>
      mathx10
      }{}
\DeclareSymbolFont{mathx}{U}{mathx}{m}{n}
\DeclareFontSubstitution{U}{mathx}{m}{n}
\DeclareMathAccent{\widecheck}{0}{mathx}{"71}
\DeclareMathAccent{\wideparen}{0}{mathx}{"75}

\RequirePackage{filecontents}        

\usepackage[numbers]{natbib}
\usepackage[colorlinks]{hyperref}    
\usepackage{cleveref}
\hypersetup{colorlinks,
    linkcolor={red!50!black},
    citecolor={blue!50!black},
    urlcolor={blue!80!black}
}

\usepackage{graphicx}
\usepackage[font=small,labelfont=bf]{caption}
\usepackage{subcaption}
\usepackage{float}
\usepackage{tikz}
\usetikzlibrary{decorations.markings,calc}
\def\ladderwidth{2.8mm}
\def\ladderstep{3.8mm}
\tikzset{
  ladder/.style={decorate,decoration={
      markings,
      mark=between positions {1/#1/2} and {-1/#1/2} step {1/#1} with {
        \draw[thick,black]
        (0,-\ladderwidth) -- (0,\ladderwidth)
        (\ladderstep,-\ladderwidth) -- (-\ladderstep,-\ladderwidth)
        (\ladderstep,\ladderwidth) -- (-\ladderstep,\ladderwidth);
      }
    },
  },
  XOR/.style={draw,circle,append after command={
      [shorten >=\pgflinewidth, shorten <=\pgflinewidth,]
      (\tikzlastnode.north) edge (\tikzlastnode.south)
      (\tikzlastnode.east) edge (\tikzlastnode.west)
      }
    },
  mmt/.style = {regular polygon, regular polygon sides=3,
            draw, fill=white,
            inner sep=0.9mm, outer sep=0mm,
            shape border rotate=-90},
  bigmmt/.style = {regular polygon, regular polygon sides=3,
            draw, fill=white,
            inner sep=2mm, outer sep=0mm,
            shape border rotate=-90},
  operator/.style = {draw,fill=white,minimum size=1.5em},
  phase/.style = {fill,shape=circle,minimum size=5pt,inner sep=0pt},
  not/.style = {shape=circle,minimum size=5pt,inner sep=0pt},
  surround/.style = {fill=blue!10,thick,draw=black,rounded corners=2mm},
}
\usetikzlibrary{backgrounds,fit,shapes,decorations.pathreplacing,patterns,arrows}
\usepackage{anyfontsize}

\usepackage{algorithm}
\usepackage{algpseudocode}

\usepackage[margin=1in]{geometry}
\hypersetup{pdfpagemode=UseNone}
\usepackage{pdfsync}

\def\field{\mathbb{F}}

\def\union{\cup}

\def\to{\rightarrow}

\newcommand{\wt}[1]{\operatorname{wt}\left( #1 \right)}

\renewcommand{\dim}[1]{\operatorname{dim}\left( #1 \right)}

\def\dsl{\llbracket}
\def\dsr{\rrbracket}

\def\C{\mathcal{C}}

\def\F{\mathcal{F}}
\def\G{\mathcal{G}}

\def\Q{\mathcal{Q}}

\def\h{\operatorname{H}}

\def\Ehat{\widehat{E}}

\newtheorem{theorem}{Theorem}
\newtheorem*{theorem*}{Theorem}
\newtheorem{lemma}[theorem]{Lemma}

\usepackage{thmtools}
\usepackage{thm-restate}

\def\path{\operatorname{P}}

\newcommand{\sansserif}[1]{%
  \ifmmode
  \mathsf{#1}%
  \else
   \textsf{#1}%
  \fi
}

\renewcommand{\int}[2]{\Delta_{#1} \G'_{#2}}

\usepackage{ dsfont }

\def\flip{\texttt{flip}}
\def\ssflip{\texttt{small-set-flip}}
\def\plog{p_{\text{Log}}}
\def\qlog{q_{\text{Log}}}
\usepackage{mathtools}
\DeclarePairedDelimiter{\ceil}{\lceil}{\rceil}
\usepackage{authblk}

\newlength\curvewidth

\title{Numerical study of hypergraph product codes}
\author[1]{Antoine Grospellier}
\author[2]{Anirudh Krishna}
\affil[1]{
  Inria, Team SECRET\\
  2 Rue Simone IFF, CS 42112\\
  75589 Paris Cedex 12, France
  }
\affil[2]{
	Universit\'e de Sherbrooke\\
	2500 Boulevard de l'Universit\'e\\
	Sherbrooke, QC J1K 2R1, Canada
}
\date{}

\begin{document}
\maketitle
\begin{abstract}
Hypergraph product codes introduced by Tillich and Z\'emor are a class of quantum LDPC codes with constant rate and distance scaling with the square-root of the block size.
Quantum expander codes, a subclass of these codes, can be decoded using the linear time small-set-flip algorithm of Leverrier, Tillich and Z\'emor.
In this paper, we numerically estimate the performance for the hypergraph product codes under independent bit and phase flip noise.
We focus on two families of hypergraph product codes.
The first family has rate $1/61 \sim 1.6\%$, has qubits of weight $10$ or $12$ and stabilizers of weight $11$.
We report a threshold near $4.6\%$ for the small-set-flip decoder.
We also show that for similar rate, the performance of the hypergraph product is better than the performance of the toric code as soon as we deal with more than $500$ logical qubits and that for $3600$ logical qubits, the logical error rate for the hypergraph product code is several orders of magnitude smaller.
The second family has rate $0.2$, qubits of  weight $10$ and $20$ and stabilizers of weight $15$.
We report a threshold near $2\%$ for the small-set-flip decoder.
\end{abstract}


\section{Introduction}
\setlength{\curvewidth}{0.97\textwidth}

It is imperative to make quantum circuits fault tolerant en route to building a scalable quantum computer.
The threshold theorem \cite{aharonov1997fault,kitaev1997quantum,knill1998resilient} guarantees that it will be possible to do so using quantum error correcting codes which encode information redundantly.
This redundancy serves as a buffer against errors but we need to be mindful of the trade-offs involved as the number of qubits we can control in the laboratory is limited.
The \emph{overhead}, defined as the ratio between the number of qubits in a fault-tolerant implementation  of a quantum circuit to the number of qubits in an ideal, noise-free environment is a figure of merit to quantify this trade-off.

Gottesman showed in \cite{gottesman2014fault} that a certain class of quantum error correcting codes called quantum low density parity check (abbrev. LDPC) codes could offer significant benefits in this regard.
These are code families $\C_n = \{\dsl n,k,d \dsr\}_n$ for which the number of qubits a stabilizer acts on remains constant with increasing block size $n$, and the number of stabilizers that a qubit is involved in also remains constant with $n$.
Such codes are ubiquitous in classical coding theory with theoretical and practical uses.
In the quantum case, we expect these codes to be useful because constant weight stabilizers in turn mean that syndrome extraction circuits will only require a constant number of ancilla qubits if we use Shor's technique for syndrome extraction \cite{shor1996fault}.
Gottesman proved that we can construct circuits with constant space overhead if we had quantum LDPC code families such that $k = \Theta(n)$ with an efficient decoding algorithm robust against noisy syndrome measurements.
This means that if we considered an ideal circuit that processes $m$ qubits, then its fault-tolerant counterpart will require $\Theta(m)$ qubits.
This result is asymptotic in nature and to ascertain its practical consequences, it would be useful to have estimates of the constants involved.

Gottesman advertised two families of codes -- hyperbolic codes (of the two- and four- dimensional varieties) \cite{freedman2002z2, guth2014quantum, londe2017golden,breuckmann2016constructions,breuckmann2017hyperbolic,conrad2018small} and hypergraph product codes \cite{tillich2014quantum,leverrier2015quantum,fawzi2018efficient,fawzi2018constant}.
It is still an open question whether or not LDPC codes exist whose distance scales linearly in the block size.
The distance of the class of $2$D hyperbolic codes is bounded by $O(\log(n))$ \cite{delfosse2013tradeoffs} whereas the distance of $4$D hyperbolic codes can scale as $n^{\epsilon}$ for $\epsilon < 0.3$ \cite{guth2014quantum}.
Hypergraph product codes possess a dimension that scales linearly with $n$, and the distance of these codes is $\Theta(\sqrt{n})$, comparable to the toric code.
Regarding decoders, the $2$D hyperbolic codes, like the toric code, utilize minimum weight matching which require $O(n^3)$ time to run for a code with $n$ qubits \cite{edmonds1965paths}.
Although this is not a problem for small codes, it eventually becomes an issue as the code grows larger.
The toric code itself does possess a parallelized decoder that runs in $O(1)$ time given constant computational resources per unit area, which works even in the fault-tolerant setting.

In \cite{leverrier2015quantum}, Leverrier et al. have shown the existence of a linear time decoder for the hypergraph product codes called small-set-flip decoder and proved it corrects errors of size $O(\sqrt{n})$ in an adversarial setting.
In \cite{fawzi2018efficient} Fawzi et al. showed that the small-set-flip decoder corrects with high probability a constant fraction of random errors in the case of ideal syndromes and in \cite{fawzi2018constant}, they made these results fault tolerant showing that this decoder is robust to syndrome noise as well.
To be precise, they showed that the small-set-flip algorithm is a single-shot decoder and that in the presence of syndrome noise, the number of residual qubit errors on the state after decoding is proportional to the number of syndrome errors.
As a consequence, hypergraph product codes can be used in Gottesman's framework to achieve fault-tolerance with constant space overhead.
Furthermore they showed analytically that this decoder has a threshold, but before this work no numerical estimate was known except an analytical lower bound of $2.7 \times 10^{-16}$ in \cite{fawzi2018efficient}.
This work ameliorates this situation by providing some numerical estimates for the performance of hypergraph product codes subject to simple noise models.
This can be contrasted to some recent work due to Kovalev et al. \cite{kovalev2018numerical}, who showed that certain hypergraph product codes achieve a threshold several orders of magnitude better than this analytical lower bound (approximately $7 \times 10^{-2}$).
It is important to note that this result is only an upper bound on what is achievable as Kovalev et al. circumvent the decoding process entirely.
Instead they indirectly estimate the probability of having a logical failure using a statistical mechanical mapping.

\textbf{Results:} In this paper, we subject hypergraph product codes to independent $X-Z$ errors and decode using the small-set-flip algorithm.
We use two families of codes obtained by forming the hypergraph product of randomly generated biregular graphs of degree $(5,6)$ and $(5,10)$ providing codes with rates $1/61 \sim 1.61\%$ and $1/5 = 20\%$ respectively.
In this setting we show that the threshold for the $(5,6)$ codes is approximately $4.6\%$ and the threshold for the $(5,10)$ codes is approximately $2\%$ albeit with some qualifications.
When we plot the logical error rates exhibit some unusual behavior as a function of the code size: they only meet when the logical error rate is very close to $1$.
However when the physical error rate is below these values, we observe typical sub-threshold behaviour, namely increasing the block length decreases the logical error rate.
To benchmark their performance, we compare the $(5,6)$ codes with the multiple copies of the toric code, chosen to match the rate as closely as possible $L = 8$.
It appears that once we exceed $500$ logical qubits the logical error rate of the $(5,6)$ hypergraph product codes is smaller than the logical error rate of the corresponding $250$ toric code copies.
Moreover increasing the number of logical qubits is beneficial to the hypergraph product and for $3600$ logical qubits, the logical error rate for the hypergraph product is $7 \times 10^{-6}$ and near $1.9 \times 10^{-2}$ for the toric code.

In section \ref{sec:background} we briefly review some material on classical and quantum expander codes and their respective decoding algorithms, $\flip$ and $\ssflip$.
We then proceed in section \ref{sec:results} to describe the results of our numerical simulations.

\section{Background}
\label{sec:background}
\subsection{Classical codes}
Consider a classical code family $\{\C_i\}_i$, where $\C_i = [n_i,k_i]$ is a binary linear code such that the block size $n_i \to \infty$ as $i\to \infty$.
This family is said to be LDPC if the weight of each row of the parity check matrix is at most $r$ and the weight of each column of the parity check matrix is at most $c$, for some natural numbers $r$ and $c$.
The weight of a row (or column) is the number of non-zero entries appearing in the row (or column).
In other words, the number of checks acting on any given bit and the number of bits in the support of any given check is a constant with respect to the block size.
These codes are equipped with iterative decoding algorithms (such as belief propagation) which have low time complexity and excellent performance.
Furthermore, they can be described in an intuitive manner using the factor graph associated with the classical code and for this reason these codes are also called graph codes.

A factor graph associated to a code $\C = [n,k]$ is a bipartite graph $\G(\C) = (V \union C, E)$ where one set of nodes $V$ represents the bits and the other set $C$ represents the checks in the code $\C$ respectively.
For nodes $v_i \in V$ and $c_j \in C$, where $i \in [n]$ and $j \in [m]$, we draw an edge between $v_i$ and $c_j$ if the $i$-th variable node is in the support of the $j$-th check.
Equivalently, if $\h$ denotes the parity check matrix of the code $\C$, we draw an edge between the nodes $v_i$ and $c_j$ if and only if $\h(i,j) = 1$.
It follows that a code $\C$ is LDPC if the associated factor graph is biregular, i.e. nodes in $V$ have degree $\Delta_V$ and nodes in $C$ have degree $\Delta_C$.

Of particularly interest are expander codes, codes whose factor graph corresponds to an expander graph.
Let $\G = (V \union C, E)$ be a bipartite factor graph such that $|V| = n$ and $|C| = m$ such that $n \geq m$.
The graph $\G$ is said to be $(\gamma_V,\delta_V)$-left-expanding if for $S \subseteq V$,
\begin{align}
	|S| \leq \gamma_V n \implies |\Gamma(S)| \geq (1-\delta_V)\Delta_V|S|~.
\end{align}
Similarly, the graph is $(\gamma_C,\delta_C)$-right-expanding if for $T \subseteq C$,
\begin{align}
	|T| \leq \gamma_C m \implies |\Gamma(T)| \geq (1-\delta_C)\Delta_C|T|~.
\end{align}
It is a \emph{bipartite} expander if it is both left and right expanding.

In their seminal paper, Sipser and Spielman \cite{sipser1994expander} studied expander codes and devised an elegant algorithm called $\flip$ to decode them.
They showed that if the factor graph is a left expander such that $\delta < 1/4$, then the $\flip$ algorithm is guaranteed to correct errors whose weight scales linearly with the block size of the code.
Furthermore, it does so in time scaling linearly with the block of the code.

$\flip$ is a deceptively simple algorithm and it is remarkable that it works.
We describe it here as it forms the basis for the quantum case decoding algorithm $\ssflip$.
Let $w \in \C$ be a codeword and $y$ be the corrupted word we receive upon transmitting $x$ through a noisy channel.
With each variable node $v_i$ in the factor graph, $i \in [n]$, we associate the value $y_i$.
With each check node $c_j$ in the factor graph, $j \in [m]$, we associate the value $x_j = \sum_{i:v_i \in \Gamma(c_j)} y_i$, where the sum is performed modulo $2$.
We use $\Gamma(c_j)$ to denote the neighborhood of the node $c_j$ in the graph $\G$.
This is merely the parity associated with the corresponding check $c_j$.
We shall say that a check node $c_j$ is unsatisfied if its parity is $1$ and satisfied otherwise.
Note that if $y \in \C$ is a codeword, then all the checks $c_j$, $j \in [m]$, must be satisfied.
Informally, $\flip$ searches for a variable node that is connected to more unsatisfied neighbors than it is to satisfied, and flips the corresponding bit.
It is stated formally below in algorithm \ref{alg:flip}.

\begin{algorithm}
	\begin{algorithmic}
		\\\hrulefill
		\State \textbf{Input:} Corrupted word $y$
		\State\textbf{Output:} Deduced error $\Ehat$ if the algorithm converges and FAIL otherwise.
	    \\\hrulefill
		\State \textbf{Algorithm:}
		\State Initialize $w \leftarrow y$. \Comment{Iteratively maintain $w$}
		\While{$\exists v_i : \sum_{j: c_j \in \Gamma(v_i)} x_j \geq \ceil{\deg(v_i)/2}$} \Comment{$\deg(v_i)$ is the degree of $v_i \in V$}
		\State $w_i \to \overline{w_i}$ \Comment{Flip the $i^\text{th}$ bit}
		\EndWhile
    \State \Return $\Ehat = y + w$ if the syndrome of $y+w$ is zero and FAIL otherwise.
	\end{algorithmic}
	\caption{\texttt{flip}}
	\label{alg:flip}		
\end{algorithm}

The number of unsatisfied checks is monotonically decreasing and therefore it is evident that the algorithm terminates in a number of steps lesser than or equal to $m$.
This implies that the algorithm terminates in linear time.
For a detailed analysis of this algorithm, we point the interested reader to the original paper by Sipser and Spielman \cite{sipser1994expander}.

\subsection{Quantum codes}
CSS quantum codes are quantum error correcting codes that only contain stabilizers each of whose elements are all $X$ or all $Z$.
They are composed of two binary linear codes $\C_Z = [n,k_1,d_1]$ and $\C_X = [n,k_2,d_2]$ such that $\C_Z^{\perp} \leq \C_X \Leftrightarrow \C_X^{\perp} \leq \C_Z$.

To construct the code,
\begin{enumerate}
  \item map the $i^{th}$ row of $\h^Z$ to $Z$ stabilizer generator $S_{i}^Z$ by mapping $1$'s to $Z$ and $0$'s to identity; and
  \item map the $j^{th}$ row of $\h^X$ to $X$ stabilizer generator $S_{j}^X$ by mapping $1$'s to $X$ and $0$'s to identity.
\end{enumerate}

The codewords correspond to cosets of $\C_Z/\C_X^{\perp}$ and hence the code dimension is $k := \dim{\C_Z/\C_X^{\perp}} = \dim{\C_X/\C_Z^{\perp}}$.
The distance is expressed as $d = \min\{d_X,d_Z\}$ where
\begin{align*}
  d_X = \min_{e \in \C_Z \setminus \C_X^{\perp}} \wt{e}
    \qquad\qquad
  d_Z = \min_{f \in \C_X \setminus \C_Z^{\perp}} \wt{f}~.
\end{align*}

A hypergraph product is a framework to construct quantum LDPC codes using two classical codes \cite{tillich2014quantum}.
The construction ensures that we have the appropriate commutation relations between the $X$ and $Z$ stabilizers without resorting to topology.
Let $\G_1$ and $\G_2$ be two bipartite graphs, i.e. for $i \in \{1,2\}$, $\G_i = (V_i \union C_i, E)$.
We denote by $n_i := |V_i|$ and $m_i := |C_i|$ the size of the sets $V_i$ and $C_i$ respectively for $i \in \{1,2\}$.

These graphs define two pairs of codes depending on which set defines the variable nodes and which set defines the check nodes.
The graph $\G_1$ ($\G_2$ resp.) defines the code $\C_1 = [n_1,k_1,d_1]$ ($\C_2 = [n_2,k_2,d_2]$ resp.) when nodes in $V_1$ ($V_2$ resp.) are interpreted as variable nodes and nodes $C_1$ ($C_2$ resp.) are represented as checks.
Note that $m_i \geq n_i - k_i$ as some of the checks could be redundant.
Similarly, these graphs serve to define codes $\C_1^T = [m_1, k_1^T, d_1^T]$ ($\C_2^T = [m_2, k_2^T, d_2^T]$ resp.) if $C_1$ ($C_2$ resp.) represents variable nodes and $V_1$ ($V_2$ resp.) the check nodes.
Equivalently, we can define these codes algebraically.
We say that the code $\C_i$ is the right-kernel of a parity check matrix $\h_i$ and the code $\C_i^T$ is the right-kernel of the matrix $\h_i^T$.
Of course, $k_i^T$ and $d_i^T$ are not transposes of $k_i$ and $d_i$ as these are mere scalars, but we use (and abuse) this notation to represent the corresponding parameters for the latter pair of codes.

We define a quantum code $\Q = \dsl n,k,d\dsr$ via the graph product of these two codes as follows.
The set of qubits is associated with the set $(V_1 \times V_2) \union (C_1 \times C_2)$.
The set of $Z$ stabilizers is associated with the set $(C_1 \times V_2)$ and the $X$ stabilizers with the set $(V_1 \times C_2)$.

\begin{lemma}
The code parameters of the code $\Q$ can be described in terms of the constituent classical codes as
\begin{enumerate}
	\item The block size of the code $\Q$ is $n = n_1 n_2 + m_1m_2$.
	\item The number of logical qubits is $k = k_1k_2 + k_1^T k_2^T$.
	\item The $X$ and $Z$ distance of the code are
	\begin{align*}
		d_X = \min(d_1^T, d_2) \qquad d_Z = \min(d_1, d_2^T)
	\end{align*}
	and therefore,
	\begin{align}
		d = \min(d_X,d_Z)~.
	\end{align}
\end{enumerate}
\end{lemma}
For the rest of this paper, we only consider the hypergraph product of two copies of the same graph.

Naively generalized to the quantum realm, $\flip$ performs poorly because of degeneracy.
There exist constant size errors that lead to the algorithm failing which implies that it will not work well in an adversarial setting.

Leverrier et al. \cite{leverrier2015quantum} address this issue by devising an algorithm called $\ssflip$ obtained by modifying $\flip$.
The algorithm $\ssflip$, presented in alg. \ref{alg:ssflip} below, is guaranteed to work on quantum expander codes which are the hypergraph product of bipartite expanders.

Let $\F$ denote the union of the power sets of all the $Z$ generators in the code $\Q$.
For $E \in \field_2^{n_1n_2 + m_1m_2}$, let $\sigma_X(E)$ denote the syndrome of $E$ with respect to the $X$ stabilizers.
Given the syndrome $\sigma_0$ of a $Z$ type error chain $E$, the algorithm proceeds iteratively.
In each iteration, it searches within the support of the $Z$ stabilizers for an error $F$ that reduces the syndrome of the the case of $X$ errors follows in a similar way by swapping the role of $X$ and $Z$ stabilizer generators.

\begin{algorithm}[h]
	\begin{algorithmic}
		\\\hrulefill
		\State \textbf{Input:} A syndrome $\sigma_0 \in \field_2^{n_1m_2}$.
		\State \textbf{Output:} Deduced error $\Ehat$ if algorithm converges and FAIL otherwise.
		\\\hrulefill
		\State \textbf{Algorithm:}
		\State $\Ehat = 0^{n_1n_2 + m_1m_2}$\Comment{Iteratively maintain $\Ehat$}
		\While{$\exists F \in \F : |\sigma_i| - |\sigma_i \oplus \sigma_X (F)| > 0$}                
		\State
		\begin{align*}
			&F_i = \arg \max_{F \in \F} \frac{|\sigma_i| - |\sigma_i \oplus \sigma_X(F)|}{|F|}\\
			&\Ehat_{i+1} = \Ehat_i \oplus F\\
			&\sigma_{i+1} = \sigma_i \oplus \sigma_X(F_i)\\
			&i = i+1
		\end{align*}
		\EndWhile
		\State \Return $\Ehat_i$ if $\sigma(\Ehat_i) + \sigma_0$ is zero and FAIL otherwise.
	\end{algorithmic}
	\caption{\ssflip($E$)}
	\label{alg:ssflip}
\end{algorithm}

The article \cite{leverrier2015quantum} proceeds to show that $\ssflip$ is guaranteed to work if the graphs corresponding to classical codes are bipartite expanders.
They prove the following theorem (theorem 2 in \cite{leverrier2015quantum})
\begin{theorem}
	\label{thm:leverrierTillichZemor}
	Let $\G = (V \union C, E)$ be a $(\Delta_V, \Delta_C)$ biregular $(\gamma_V,\delta_V,\gamma_C,\delta_C)$ bipartite expander.
	Suppose $\delta_V < 1/6$ and $\delta_C < 1/6$. Further suppose that $(\Delta_V, \Delta_C)$ are constants as $n$ and $m$ grow.
	The decoder $\ssflip$ for the quantum code $\Q$ obtained via the hypergraph product of $\G$ with itself runs in time linear in the code length $n^2 + m^2$, and it decodes any error of weight less than
	\begin{align}
		w = \frac{1}{3(1+\Delta_C)}\min(\gamma_V n, \gamma_C m)~.
	\end{align}
\end{theorem}

The followup paper by \cite{fawzi2018efficient} demonstrated that the $\ssflip$ algorithm could correct with high probability a number of random errors growing linearly with the block size of the code.
Furthermore, they showed that the hypergraph product codes equipped with $\ssflip$ exhibited a threshold.
Their analytical estimates lower bound this number by $2.7 \times 10^{-16}$.
The recent result by \cite{fawzi2018constant} showed that if we had corrupt syndrome measurements, then the $\ssflip$ algorithm was still capable of reducing the total number of errors.

\section{Numerical simulations of small-set-flip}
\label{sec:results}

In this section, we present the result of numerical simulations of the $\ssflip$ algorithm on a set of hypergraph product codes constructed using two copies of the same biregular graph.
The resulting hypergraph product codes that we obtain from the classical codes above are subject to independent bit and phase flip noise.
We assume that each qubit is independently afflicted by $X$ and $Z$ errors with some probability $p$.
This implies that the quantum codes, being CSS codes, can be decoded separately.
We study below the logical error rate $\plog$ versus the physical error rate $p$, where the quantity $\plog$ is the probability of failure for \emph{at least one} logical qubit to be corrupted.
Equivalently, $1-\plog$ is the probability that none of our logical qubits are corrupted and is called the \emph{logical success rate}.

The biregular graphs we have used below were generated randomly using the configuration model (see \cite{richardson2008modern}).
Even in the presence of multi-edges, it is possible to show that asymptotically, these graphs will have a good expansion co-efficient.
We found a correlation between the performance of $\flip$ on the classical code whose factor graph is the biregular graph and the performance of $\ssflip$ on the resulting quantum codes.
Thus we benchmarked the performance of these graphs using $\flip$ on the corresponding classical codes and picked the best among them to use as quantum codes.

We present two classes of codes constructed using the hypergraph product of a family of $(5,6)$- and $(5,10)$- biregular bipartite graphs.
Throughout the paper, we refer to this code family using the degrees of the classical factor graph as the $(5,6)$ and $(5,10)$ codes.
Among the codes that we tested, the $(5,6)$ family appears to have the best performance; $\ssflip$ does not appear to work for graphs with smaller degrees.
Note that the result of \cite{leverrier2015quantum} required a graph whose left and right degrees were at least $7$ to guarantee good performance; our simulations indicate that we can do better with smaller degrees.
The resulting quantum codes have qubits whose degrees are either $10$ or $12$ whereas the weight of both the $X$ and $Z$ stabilizers is $11$.
These codes have a fairly low, yet constant rate of $1/61 \approx 0.016$.
The logical error rates have been plotted in fig. \ref{fig:56} below.
The rate for the $(5,10)$ family is considerably higher at $0.2$.
The qubits of these codes have degree $10$ or $20$ and the weight of the stabilizers is $15$.
The logical error rates have plotted in fig. \ref{fig:510} below.

In all plots, the integers $n$ and $m$ are respectively the number of bits and check-nodes for the classical error correcting codes used to construct the $\dsl n^2 + m^2, (n-m)^2\dsr$ quantum LDPC codes.
The error bars for all plots represent the $99\%$ confidence intervals, i.e $\approx 2.58$ standard deviation.

\begin{figure}[!h]
	\centering
        \includegraphics[width=\textwidth]{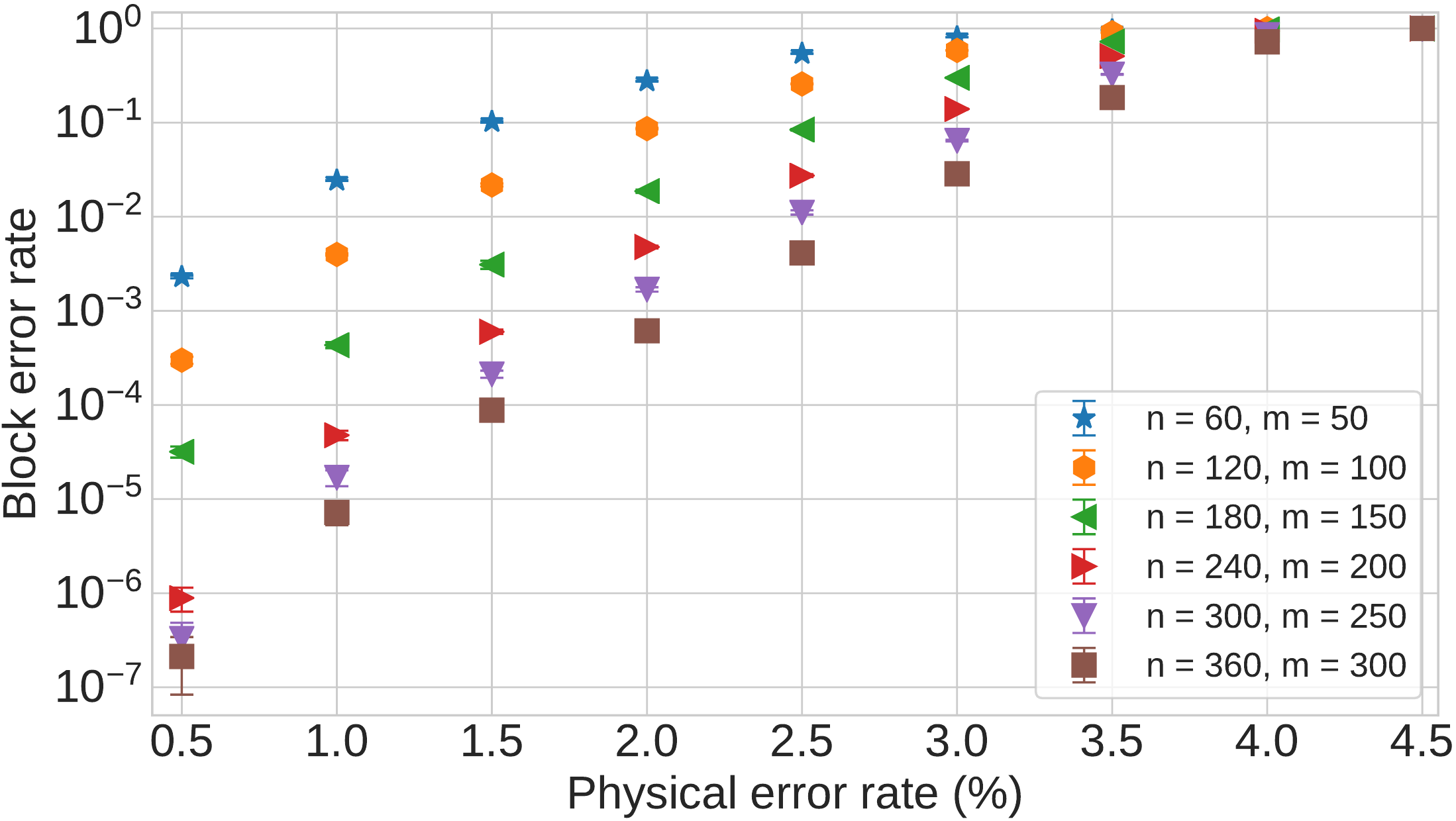}
	\caption{Block logical error rates for hypergraph product of $(5,6)$-biregular graph with itself as a function of the physical error rate $p$.
	}
	\label{fig:56}
\end{figure}
\begin{figure}[!h]
	\centering
        \includegraphics[width=\textwidth]{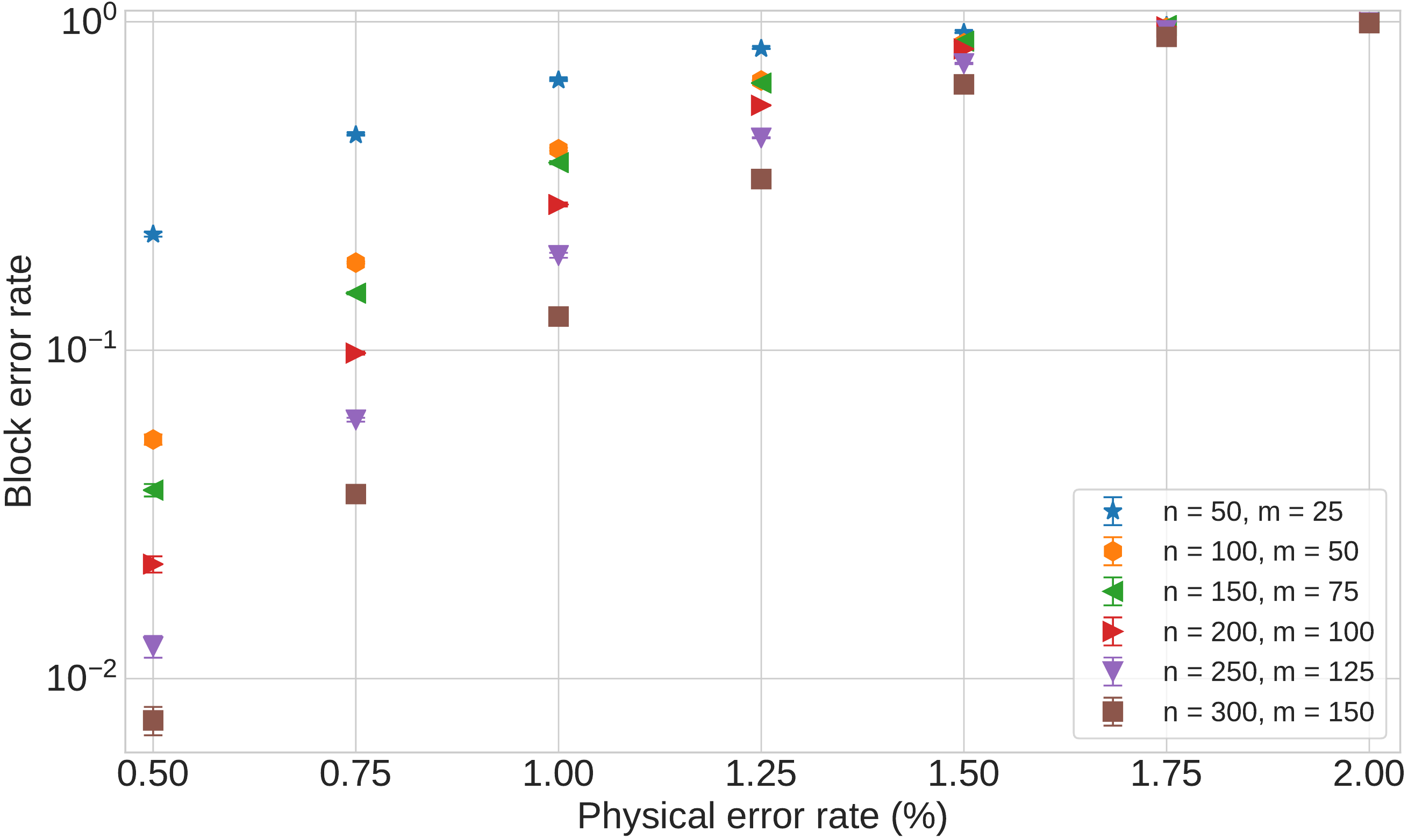}
	\caption{Block logical error rates for hypergraph product of $(5,10)$-biregular graph with itself as a function of the physical error rate $p$.}
	\label{fig:510}
\end{figure}

All code families we tested exhibit some unusual behavior in that the curves for the logical error rate of different block sizes cross only when the logical error rate is close to $1$.
Although this makes the notion of a threshold ambiguous, we do find that for all families, there exists a physical noise rate below which the logical error rate is lower for larger block lengths.
Using this as an indication of sub-threshold behavior, the $(5,6)$ family has a threshold of roughly $4.6\%$ and the $(5,10)$ family has a threshold of roughly $2\%$.

To benchmark the performance of these codes, we compare our results with the toric code.
Assume that we have a hypergraph product code with $k$ logical qubits that exhibits a logical failure of $\plog(p)$ at noise rate $p$.

Further let $\qlog(p)$ denote the logical error rate of the toric code of length $L = 8$ at noise rate $p$.
We compare the logical failure probability of the block when we encode $k$ qubits using $k/2$ copies of the toric code versus the hypergraph product code.
Let $\qlog(p)$ denote the logical failure rate of the toric code of length $L = 8$ subject to independent bit and phase flip noise at noise rate $p$, and decoded using minimum weight matching.
The side length $L = 8$ is chosen such that $2/L^2$ is as close as possible to the rate $k/n$ of the hypergraph product code.
The block is said to have a logical failure if at least one logical qubit suffers a logical error.
Specifically, we compare $(1-\qlog)^\frac{k}{2}$ and $(1-\plog)$.
For the case of the $(5,6)$ code, we find that the hypergraph product codes of the same rate perform better after roughly $500$ logical qubits.
This has been plotted in fig. \ref{fig:5,6_VS_toric} below.

\begin{figure}[!h]
  \centering
  \includegraphics[width=\curvewidth]{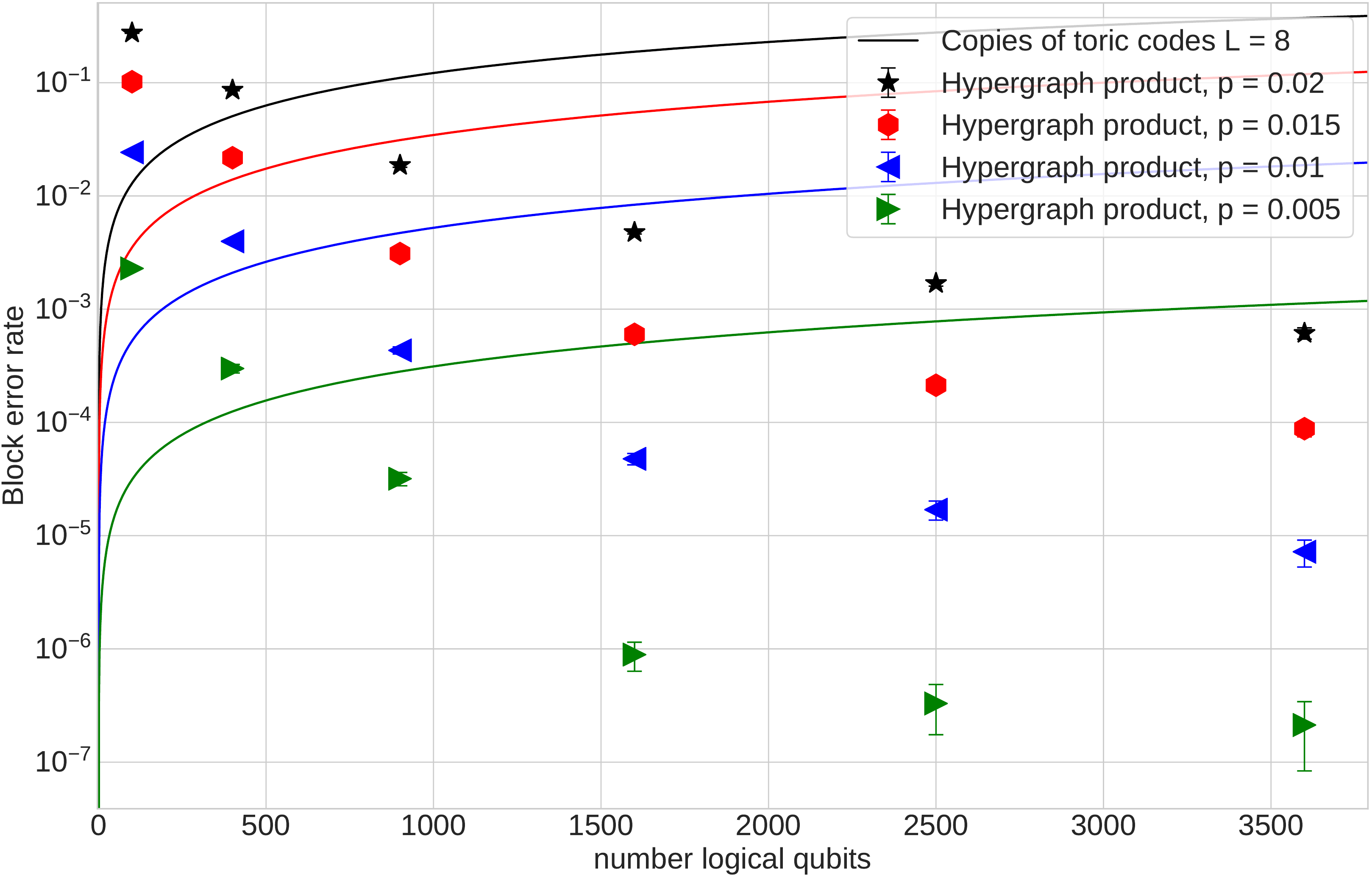}
  \caption{Logical error rate versus $k$ the number of logical qubits for hypergraph product of $(5,6)$-biregular graph and for $k/2$ toric code copies.
  }
	\label{fig:5,6_VS_toric}
\end{figure}

Note that this estimate is not meant to be indicative of the dimension of the logical space for when all hypergraph product codes become advantageous with respect to the toric code.
In fact for codes with a higher rate, we expect hypergraph product codes that encode $k$ qubits to be better than $k/2$ copies of the toric code for much smaller values of $k$.
For the case of the $(5,10)$ code, the appropriate toric code to compare with is one for which the length $L=3$.
We have not included this comparison because it is unclear what we can infer for such a small code.

\section{Conclusion}
We studied the performance of the small-set-flip algorithm on instances of hypergraph product codes numerically.
We presented the results of our simulations of the algorithm on families created by the product of classical codes whose factor graphs are $(5,6)$-biregular bipartite graphs and $(5,10)$-biregular bipartite graphs.
For noiseless error correction, i.e. assuming that syndrome measurements are ideal, it appears that these codes have a threshold of roughly $4.6\%$ and $2\%$ when subject to independent bit and phase flip noise.
We then compared the logical error rate of the $(5,6)$ codes to the toric code and estimated that the hypergraph product code is beneficial after a block size corresponding to roughly $500$ logical qubits.
As we increase the block size, we find that our logical failure probability improves by several orders of magnitude.
These comparisons indicate that to achieve a target logical error probability, these codes could offer significant savings in overhead for large block sizes.

These results are promising and indicate that numerical simulations on hypergraph product codes warrant further attention.
Future research could study these codes with detailed noise models, including syndrome noise or even circuit level noise.
Moreover, it is known that even in the classical case, the $\flip$ algorithm required large block sizes before it performed well \cite{richardson2008modern}.
LDPC codes have become ubiquitous because of iterative decoding algorithms such as belief propagation which improved the performance of these codes significantly.
It would be interesting to know whether there are better decoding algorithms than $\ssflip$ for quantum LDPC codes.

\section{Acknowledgments}
We would like to thank David Poulin and Jean-Pierre Tillich for discussions and feedback on this project.
We would also like to thank Omar Fawzi, Anthony Leverrier and Vivien Londe for useful comments.
AG would like to thank the INTRIQ organization for an internship grant that facilitated his visit to the Universit\'e de Sherbrooke and to the group of David Poulin for their hospitality.
AG acknowledges support from the ANR through the QuantERA project QCDA.
AK would like to thank the MITACS organization for the Globalink award which partially supported his visit to Inria, Paris, where part of this research was done.
AK also acknowledges the support of the FRQNT for the B2X scholarship.
Computations were made on the supercomputers managed by Calcul Qu\'ebec and Compute Canada.
The operation of these supercomputers is funded by the Canada Foundation for Innovation (CFI), the minist\`ere de l'\'Economie, de la science et de l'innovation du Qu\'ebec (MESI) and the Fonds de recherche du Qu\'ebec - Nature et technologies (FRQ-NT).

\bibliographystyle{unsrtabbrev}
\bibliography{references}

\end{document}